# DeepHammer: Depleting the Intelligence of Deep Neural Networks through Targeted Chain of Bit Flips

Fan Yao
*University of Central Florida*
fan.yao@ucf.edu

Adnan Siraj Rakin  Deliang Fan
*Arizona State University*
asrakin@asu.edu  dfan@asu.edu

## Abstract

Security of machine learning is increasingly becoming a major concern due to the ubiquitous deployment of deep learning in many security-sensitive domains. Many prior studies have shown external attacks such as adversarial examples that tamper the integrity of DNNs using maliciously crafted inputs. However, the security implication of *internal threats* (i.e., hardware vulnerability) to DNN models has not yet been well understood.

In this paper, we demonstrate the first hardware-based attack on *quantized deep neural networks*–**DeepHammer**–that deterministically induces bit flips in model weights to compromise DNN inference by exploiting the rowhammer vulnerability. DeepHammer performs aggressive bit search in the DNN model to identify the most vulnerable weight bits that are flippable under system constraints. To trigger deterministic bit flips across multiple pages within reasonable amount of time, we develop novel system-level techniques that enable fast deployment of victim pages, memory-efficient rowhammering and precise flipping of targeted bits. DeepHammer can deliberately degrade the inference accuracy of the victim DNN system to a level that is only as good as **random guess**, thus *completely depleting the intelligence of targeted DNN systems*. We systematically demonstrate our attacks on real systems against 12 DNN architectures with 4 different datasets and different application domains. Our evaluation shows that **DeepHammer** is able to successfully tamper DNN inference behavior at run-time within a few minutes. We further discuss several mitigation techniques from both algorithm and system levels to protect DNNs against such attacks. Our work highlights the need to incorporate security mechanisms in future deep learning system to enhance the robustness of DNN against hardware-based deterministic fault injections.

## 1 Introduction

Machine learning services are rapidly gaining popularity in several computing domains due to the tremendous advancements of deep learning in recent years. Because of the unparalleled performance, deep neural networks (DNNs) are widely used nowadays in many decision-making tasks including access control [11, 17, 19, 29, 63], malware detection [69], medical diagnostics [49] and autonomous driving [8, 55]. With such ever-increasing interaction between intelligent agents and human activities that are *security and safety critical*, maintaining security objectives (e.g., confidentiality, integrity) is the first-order design consideration for DNN systems [51].

While considerable attention has been focused on protecting DNN against input-based external adversaries (e.g., adversarial examples and data poisoning attacks [3, 7, 42, 62]), we note that *internal adversaries* that leverage vulnerability of commercial-off-the-shelf hardware is a rapidly rising security concern [6]. Recent development of fault injection threats (e.g., rowhammer attack [27]) can successfully compromise the integrity of data belonging to a victim process, which can lead to catastrophic system breach such as privilege escalation [48]. These hardware-based attacks are extremely worrisome since they are capable of directly tampering the *internal state* of a target system. In light of the power of the hardware-based threat, we note that understanding its security implication in deep learning systems is imperative.

Recently Hong et al. [21] have shown that single-bit corruptions in DNN model parameters can considerably degrade the inference accuracy of several DNN models. Their attack study is performed on full-precision (i.e., floating-point numbers) DNN models where a single bit flip in the exponent field (i.e., the most-significant bit) of a parameter can result in orders of magnitude change in the parameter value. Note that *quantized deep neural networks* [23], on the other hand, are more robust to single-bit corruption. This is because model quantization replaces full-precision model parameters with low bit-width integer or even binary representation, which significantly limits the magnitude of possible parameter value range [22, 70]. Our initial investigation aligns with this observation in [21] that single bit flip in quantized model weights does not introduce any observable accuracy loss for 99% of the time. Due to the impressive improvement in energy efficiency, memory footprints and storage, model quantization is now the *de facto* optimization in deep neural networks [71]. Yet it remains uncertain whether a successful rowhammer attack on quantized neural network is possible.

In this paper, we present a new class of DNN model fault injection attack called *DeepHammer* that targets quantized deep neural networks. DeepHammer flips a small set of targeted bits to precisely degrade the prediction accuracy of the target model to the level of random guess. We systemically characterize how bit flips of model parameters can influence the accuracy of a well-trained quantized deep neural networks. Our study focuses on model weights as these are the major components of DNN model with most substantial impact on DNN performance. Our findings indicate that to carry out a successful fault injection attack, multiple bit flips spanning many layers of the model are required. This can be extremely more difficult as compared to previous rowhammer exploits due to major algorithmic and system-level challenges.

*The first challenge* involves designing an effective bit search algorithm that understands system constraints and minimizes the number of bit flips at the same time. This is necessary because flipping a certain combination of bits may not be possible if the DRAM profile of flippable locations does not allow. Furthermore, even if multiple bit flips are attainable, the attack is unlikely to succeed if the targeted bits in the model are simply numerous. In other words, the targeted bits in model weights should be as few as possible. *The second challenge* lies in developing an efficient rowhammer attack that could successfully flip multiple bits within a reasonable exploitation window. We note that even with very small number of bits to flip, the exploitation can still be unreasonably long. In fact, *Gruss et al.* have recently shown that a single bit flip in the victim's memory can take a few days to accomplish [13]. As the disturbance errors in DRAM are transient, shortening the exploitation window for multi-bit flips is critical since the flipped bits generally do not persist after a memory reset or system reboot.

To tackle the first challenge, we propose a *bit search method* to perform bit-wise gradient ranking combined with progressive search to find the least amount of vulnerable bits that are most influential in the targeted model. Since the generated bit locations may not be empirically flipped, we implement a flip-aware search technique that takes into account several *system constraints* relating to victim's memory layout and target DRAM bit flip profile. The bit search process generates a chain of targeted bits and ensures that these bits can be physically flipped in the target machine. If the chain of bits are *all flipped*, the attacker could eventually compromise the target model. Importantly, we find that the bit chain is *not unique* for each model, and our search algorithm can potentially generate many distinct bit chains to implement the attack.

DeepHammer addresses the second challenge by developing a new rowhammer attack framework with several novel enhancement techniques. Our attack implementation enables hammering of a sequence of target bits. Additionally, we observe that to achieve the desired accuracy loss, attacker needs to *precisely* flip the desired bits. That is, flipping extra bits besides the targeted chain of bits surprisingly *alleviates* accuracy loss. Therefore, a native approach of probabilistic row hammering would not suffice. DeepHammer incorporates three advanced attack techniques to enable *fast* and *precise* row hammering: 1) advanced memory massaging that takes advantage of per-cpu free page list for swift vulnerable page relocation. This allows an order of magnitude faster page relocation as compared to the state-of-the-art memory waylaying approach [21], 2) precise double-sided rowhammering using *targeted column-page-stripe* data pattern to guarantee exact (i.e., no more and no less) bit flips in the victim DNN model; 3) online memory re-templating to update obsolete bit flip profile. The combined rowhammer attack techniques can successfully induce bit errors in the target locations that lead to the attacker-desired accuracy loss.

In summary, we make the following key contributions:

- We highlight that multiple deterministic bit flips are required to attack quantized DNNs. An efficient flip-aware bit search technique is proposed that identifies the most vulnerable model bits to flip. The search algorithm models system constraints to ensure that the identified bits can be flipped empirically.
- We develop a new rowhammer attack framework tailored for inducing bit flips in DNN models. To achieve the desired accuracy loss and have a reasonable exploitation window, our attack employs several novel enhancement techniques to enable fast and precise bit flips.
- We implement an end-to-end DeepHammer attack by putting the aforementioned techniques together. We evaluate our attacks on 12 DNN architectures with 4 datasets spanning image classification and speed recognition domains. The results show that the attacker only needs to flip between 2-24 bits (out of millions of model weight parameters) to completely compromise the DNN model. DeepHammer is able to successfully attack the targeted chain of bits in minutes.
- We evaluate the effectiveness of DeepHammer with single-side rowhammer method and using DRAM configurations with a wide spectrum of bit flip vulnerability levels. Our results show that DeepHammer can still succeed under most of such configurations.
- We investigate and evaluate mitigation techniques to systematically protect targeted, multi-bit fault injection attacks for quantized neural networks via DeepHammer. Our work calls for algorithmic and system-level techniques to enhance the robustness of deep learning systems against hardware-based threats.

## 2 Background

In this section, we present the background related to the proposed work in this paper including deep neural networks, DRAM organizations, and rowhammer attacks.

**Deep neural networks.** DNNs have been shown to be very efficient in many modern machine learning tasks. A typical deep neural network consists of a multi-layered structure including input layers, many hidden layers and one output layer. Essentially, deep neural networks are configured to approximate a function through a training process using a labeled data set. Training a DNN model involves forward- and backward-propagation to tune DNN parameters (such as model weights) with the objective of minimizing prediction errors. Due to the existence of large number of parameters in DNN models and the enormous computation with respect to parameter tuning, the training procedure can be extremely time- and resource-consuming. Moreover, well-trained DNN models generally need large amount of training data that may not be always accessible. Therefore, to expedite the process of deployment, developers tend to utilize pre-trained models released by third parties (e.g., ModelZoo [1]).

Recent years, there are many advancements towards generating efficient and compact deep neural networks through various compression techniques such as network pruning and quantization [25, 71]. Notably, quantization replaces full-precision DNN models with low-width or even binarized parameters that can significantly improve the speed and power efficiency of DNN inference without adversely sacrificing accuracy [15, 23]. Consequently, model quantization techniques have been used widely in deep learning systems, especially for resource-constrained applications [14].

**Rowhammer attacks.** Rowhammer is a class of fault injection attacks that exploits DRAM disturbance errors. Specifically, it has been shown that frequent accesses on one DRAM row (i.e., activation) will lead to toggling of voltage on DRAM word-lines. This amplifies inter-cell coupling effects, leading to quicker leakage of capacitor charge [27]. If sufficient charge is leaked before the next scheduled refresh, the memory cell will eventually lose its state, and a *bit flip* is induced. By carefully selecting neighboring rows (aggressor rows) and performing frequent row activations, an adversary can manage to modify some critical bits without access permissions (e.g., kernel memory or data in other address spaces). To trigger bit flips, there are mainly three hammering techniques: 1) single-sided rowhammer manifests by accessing one row that is adjacent to the victim row [48]; 2) double-sided rowhammer alternatively accesses *two adjacent rows* to the victim row [30, 45, 48]; 3) one-location hammering that accesses only one location in one row repeatedly to attack the target row [13]. Double-sided rowhammer attack typically generates most bit flips as it introduces the strongest cross-talk effect for memory cells in the target row [27].

## 3 Threat Model and Assumptions

Our attack targets modern DNNs that are quantized where model parameters are in the form of low bit-width integer numbers (i.e., 8-bit). The adversary manages to trigger DNN model bit flips in DRAM *after* the victim models are deployed for inference. This is different from prior attacks that inject stealthy payloads to the DNN model and re-distribute it to victim users (e.g., DNN trojaning attacks [38]). We assume that the deep neural network is setup on a resource-sharing environment to offer ML inference service. Such application paradigm is becoming popular due to the prevalence of machine-learning-as-a-service (MLaaS) platforms [46].

The objective of attacker is to compromise DNN inference behavior through inducing deterministic errors in the model weights by exploiting the rowhammer vulnerability in DRAM. Our attack aims to drastically degrade the inference accuracy of the target DNN models. The attack is regarded as successful if inference accuracy is close to random guess after the exploitation. We note that while adversarial inputs [7, 42] can also influence inference accuracy, our attack is fundamentally different: While adversarial inputs only target miss-classification for specially crafted *malicious inputs*, our attack aims to degrade the overall inference accuracy for *legitimate inputs*.

We assume that the attacker is aware of the model parameters in the target deep learning systems. Particularly the model weight parameters are known to the attacker. Such assumption is legitimate due to two main reasons: 1) As training process is typically expensive, deploying machine learning service using publicly available pre-trained models is the trending practice; 2) Even for private models, it is possible for adversaries to gain knowledge of model parameters through various form of information leakage attacks (e.g, power, electromagnetic and microarchitecture side channels [2, 12, 64–68]).

The attacker is co-located with the victim DNN service, and can run user-space unprivileged processes. Additionally, it can map pages in the weight file to its own address space in read-only mode. To map virtual address to physical address, the attacker can take advantage of huge page support. If such support is not available in the system, the attacker can leverage hardware-based side channels [13] or use advanced memory massaging techniques [30]. In this work, we mainly harness double-sided rowhammer technique as it has been shown to be most effective in inducing bit flips. Double-sided rowhammer relies on a settlement of two adjacent rows to the victim row, and thus requires knowledge of DRAM addressing scheme, which could be obtained through reverse engineering [43]. We assume that proper software-level confinement policies (e.g., process isolation) are in place. We further assume that the operating system and hypervisor are benign and up-to-date, and kernel protection mechanisms (e.g., [4]) are employed to guard kernel data structures (e.g., page tables) against full system exploitation. Such assumptions are aligned with recent rowhammer exploitations (e.g., [13, 21, 30]).

## 4 DeepHammer Overview

In this section, we present an overview of our DeepHammer attack approach. The attack has two off-line steps and one on-line step. The first off-line step is memory templating phase that finds vulnerable bit offsets in a set of physical pages. In the second off-line step, DeepHammer runs a flip-aware bit search algorithm to find the minimal set of bits to target. During the online phase, DeepHammer locates the pages containing exploitable bits and trigger multiple bit flips using several advanced rowhammer techniques.

**DRAM bit flip profiling.** In order to deterministically trigger bit flips in the target DNN model, the first step is to scan the memory for bit locations that are susceptible to bit flips. This process is also called *memory templating* [45], which is typically considered an offline preparation step. For double-sided rowhammering, the attacker has to understand the physical address to row mapping scheme. We reverse-engineer the DRAM addressing schemes for several different hardware configurations using technique proposed in [43]. Since the profiling is performed in the attacker's own memory space, it does not affect the normal operation of the underlying system. The memory templating phase generates a list of physical pages (identified by page frame numbers) together with vulnerable *bit offset in page*, flip direction ($1\rightarrow0$ or $0\rightarrow1$) and the probability of observing bit flip.

**Vulnerable bit search in DNN model.** We develop a flip-aware bit search technique that takes as input the bit flip profile generated in the profiling stage. Our algorithm aims to locate the least number of bits (i.e., the least number of physical pages) to attack in order to yield the desired accuracy loss (i.e. accuracy close to random guess in this work). The proposed technique consists of two major components: *Gradient based Bit Ranking* (GBR) and *Flip-aware Bit Search* (FBS). It performs aggressive search using bit-wise gradient ranking. The search technique ranks the influence of model weight bits in the target DNN model based on gradient. It then employs flip-aware search which identifies the most vulnerable bits that are flppable and minimize the number of necessary flips. *Missing one target bit or flipping a bit at a wrong location may significantly deteriorate the attack outcome*. We found that it is extremely important to consider system constraints to guarantee the identified bits could be flipped empirically. For instance, multiple bits could map to several weight parameters in the same virtual 4KB boundary, which could make it impossible to find a satisfactory physical page. To ensure that the vulnerable bits found could be flipped through rowhammer, the algorithm searches through flippable page offsets based on the DRAM bit flip profile. To enhance the success rate of relocating the target page (that has the target bit), we further optimize the search algorithm by prioritizing model weight bits which have higher number of candidate physical locations.

**Fast and precise bits flipping using rowhammer.** The on-line exploitation phase launches rowhammer attack to flip the chain of bits identified by the bit search algorithm. The major challenge of this process is to position victim pages to the vulnerable DRAM rows. Prior studies have shown that page positioning or memory massaging is the most time-consuming step [13]. To enable fast memory massaging, our attack exploits a specific kernel data structure: `per-cpu pageset`, which is maintained by linux operating system as a fast cache for recently freed pages. The per-cpu pageset adopts *Last-In-First-Out* policy for page allocation. Our attack takes advantage of the per-cpu pageset for fast release and remap of of vulnerable physical pages. To induce precise bit flips, we apply a new data pattern called column-page-stripe to the aggressor and victim pages. Such technique allows the attacker to induce $1\rightarrow0$ and $0\rightarrow1$ flipping *simultaneously* in a single hammering iteration for targeted bits while making sure that irrelevant bits are kept changed. Moreover, we found that bit flip profile generated in the profiling stage can be obsolete after system reboot due to memory scrambling [27]. Interestingly, we observe that memory scrambling merely alternates *the direction of the flip* (e.g., from $1\rightarrow0$ to $0\rightarrow1$ ) and does not change vulnerable bit locations. Based on this observation, we propose a technique named *online memory re-templating* to swiftly correct inconsistent bit flip profile.

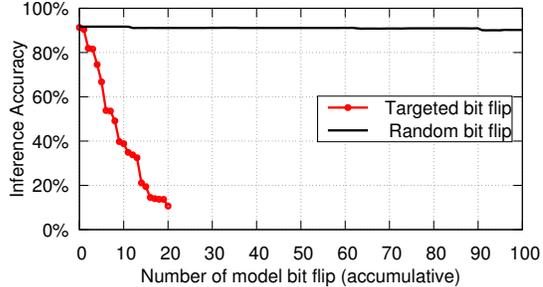

Figure 1: Randomly model bit flipping vs. targeted bit flipping for quantized ResNet-20 with CIFAR-10.

## 5 Flip-aware Vulnerable Bit Search

In this section, we first motivate the need for carefully identifying vulnerable bits in order to compromise a quantized network. We perform a robustness study of DNN models by injecting faults to model weight parameters. Figure 1 shows the changes of prediction accuracy under two bit flipping strategies for 8-bit quantized ResNet-20 using the CIFAR10 dataset [28]. As we can see, randomly flipping even 100 bits in model weights barely degrade the model accuracy to a noticeable level (i.e., less than 1%). This indicates that quantized DNNs have good tolerance against model bit flips. Note that most prior successful fault injection techniques based on rowhammer manifest by only exploiting one or very few bit flips [9, 13, 48]. Therefore, to practically carry out bit flip attack in quantized DNNs, the attackers need to find ways to identify and target the least amount of bits in models that are most vulnerable. Figure 1 further demonstrates that with our proposed targeted bit flip scheme (detailed later), attacker can considerably disrupt the inference behavior with a very small number of bit flips.

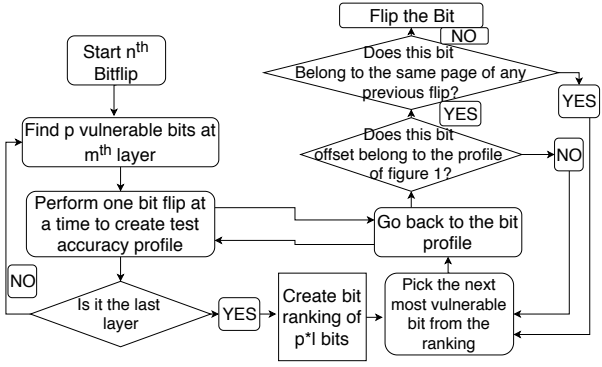

Figure 2: Overview of our proposed bit search framework.

To attack quantized DNN models, we propose an efficient flip-aware vulnerable bit search algorithm. Instead of searching all the bits of a network to generate a set of vulnerable bits, our algorithm utilizes a gradient based ranking to select top-ranked vulnerable bits [1]. The proposed method also considers the feasibility of a certain bit-flip by considering the memory layout of the model weight parameters.

In order to identify both vulnerable and flippable model bits, we first need to understand model weight storage and the corresponding memory layout. In this work, we qunatize the weights to 8-bit representations following standard quantization and representation techniques [71]. Consider a DNN model with $l$ number of layers, each layer has a weight tensor containing the weights of that particular layer. Again, each of those weights would require 8 bits memory space. Assume that the memory footprint of model weights is M, and M=$T \times 8$ bits, where $T$ is the total number of weight parameters for a particular DNN. Since weight files are loaded into memory using multiple physical pages (with typical size of 4KB), the total number of pages required for a particular DNN would be $M/4096$. Inside every page, each weight parameter has a byte offset (0-4095) and each bit has a bit offset (0-32767). Since each physical page has a deterministic DRAM mapping and the location of weak cells in DRAM modules are mostly fixed, only certain bit offsets (if any) in any physical page are vulnerable to bit flips. This profile changes across different DRAM modules (even for devices from the same vendor). Our flip-aware bit search algorithm manages to identify a certain highly vulnerable bit and attempt to find a placement of its physical page such that the targeted vulnerable bit is flippable. The algorithm optimizes the number of such flippable bits to achieve the attack goal. At a high level, our algorithm has two major steps: 1) Gradient based bit ranking which ranks the top vulnerable bits of weight parameters in a victim DNN model based on gradient; 2) Flip-aware bit search that generates *a chain of flippable bits* to target by modeling system constraints based on DRAM bit flip profile. The overall bit search framework encompasses several iterations. Each iteration interleaves the two aforementioned steps and involves identifying one model bit to flip. Our algorithm currently consider flipping only one bit for each physical page that stores model weights.

---

[1] Note that Rakin et al. [44] recently demonstrate a preliminary algorithmic work in bit-flip attack to locate vulnerable bits of DNN model. It assumes ideal scenarios where any arbitrary bit in DNN models is flippable, which is not practical in realistic settings.

**Gradient-based bit ranking (GBR):** In this step, we create a ranking of most vulnerable bits in the network based on its gradient values. Assume that the current iteration is $n$, we use $\{\hat{\mathbf{B}}_m\}_{m=1}^{l}$ to represents the original weights of the target DNN model in 2's complement form. $\hat{\mathbf{B}}_m^n$ denotes the model weights in the $n^{th}$ iteration (i.e., $n-1$ bits have already been selected and flipped). The goal is to find the $n^{th}$ bit to flip on top of the prior $n-1$ flips such that the accuracy drop is maximized in the current iteration. We find the $p$ most vulnerable bits from $\hat{\mathbf{B}}_m^n$ in $m$-th layer through gradient ranking for all the $l$ layers. With the given input $\mathbf{x}$ and target label $\mathbf{t}$, inference and back-propagation operations are performed to compute the gradients of bits w.r.t. the inference loss. Then, we select $p$ vulnerable bits that have top absolute gradient values (i.e., $\partial \mathcal{L}/\partial b$). The top-$p$ vulnerable bits can be defined as:

$$\hat{\boldsymbol{b}}_m^{n-1} = \underset{p}{\text{Top}} \left| \nabla_{\hat{\mathbf{B}}_m^{n-1}} \mathcal{L}\left(f(\boldsymbol{x}; \{\hat{\mathbf{B}}_m^{n-1}\}_{m=1}^{l}), \boldsymbol{t}\right) \right| \quad (1)$$

where $\{\text{Top}_p\}$ returns a set of bit offsets of those selected $p$ vulnerable bits, and $f(.)$ is the inference function of the DNN. By repeating the above process for all the $l$ layers, we have a candidate of $p \times l$ bits. We then evaluate the potential loss increment and accuracy degradation caused by flipping each of those vulnerable bits. The bit that causes maximum accuracy drop when flipped would be chosen in the current iteration. The corresponding loss of flipping the $i^{th}$ bit (i=1,2 ,..., p×l) in the candidate bit set–$\mathcal{L}_i^n$–can be formulated as:

$$\mathcal{L}_i^n = \mathcal{L}\left(f(\boldsymbol{x}; \{\hat{\mathbf{B}}^n\}_{i=1}^{l \times p}, \boldsymbol{t}\right) \quad (2)$$

where the only difference between $\{\hat{\mathbf{B}}^n\}$ and $\{\hat{\mathbf{B}}^{n-1}\}$ are the additional bit that is currently under test among the $p \times l$ bits, denoted as $\hat{\boldsymbol{b}}^n$. Note that, after the loss and accuracy degradation has been evaluated, GBR will continue to evaluate the next bit in the candidate. To do so, the bits flipped represented by $\hat{\boldsymbol{b}}^n$ will have to be restored back to its original state $\hat{\boldsymbol{b}}^{n-1} \in \{\hat{\mathbf{B}}^{n-1}\}$. GBR will finally generate a complete ranking of the $p \times l$ bits for the network. The information of these bits including flip direction, page number, page offset within the page, test accuracy after flipping are collected and stored.

**Flip-aware bit search (FBS):** In this step, we perform flip-aware bit search to discover a chain of bit flips that can degrade the inference accuracy to the desired level on the target hardware platform. FBS takes as input the top-ranking vulnerable bits identified by GBR. It also requires access to the DRAM bit flip profile specifying physical page frames and the page bit offsets where bit flip with certain direction (i.e., 1→0 or 0→1 ) could be induced. For the current iteration $n$, after the GBR step is complete, FBS start to iterate over the vulnerable bits in a greedy fashion by examining the bit with highest impact on test accuracy first. Specifically, it refers to the bit flip profile to check whether there is at least one available physical page (i.e., DRAM location) where the bit could be flipped [2]. That is, if both the bit offset and flip direction match, this model weight bit is considered flippable and would be inserted to the targeted bit chain. Otherwise, this bit is skipped since flipping is not possible in the victim hardware setting. The algorithm will then move on to analyze the next vulnerable bit candidate. FBS accumulatively evaluates the inference accuracy degradation due to flipping all bits in the bit chain. If the accuracy drop reaches the attack objective, the search is complete and the targeted bit chain will be collected. Otherwise, the selected bit to target in the $n^{th}$ iteration is bookmarked, and the next iteration begins with the GBR step that performs gradient ranking again. Figure 2 illustrates the overall mechanism of our bit search framework.

## 6 Fast and Precise Multi-bit Flips

By running the bit search algorithm as described in Section 5, the attacker collects one or multiple chains of bits to target in the victim DNN model. The attacker now needs to properly locate the corresponding victim pages to the vulnerable DRAM rows, and precisely induce the desired bit flips. In this section, we present three novel techniques to enable fast and precise multi-bit rowhammering. Specifically, in Section 6.1 we introduce a multi-page memory massaging techniques that exploits CPU local page cache to accurately position the target victim pages. Section 6.2 illustrates the design of our *precise hammering scheme* which ensures only the desired bits are flipped. We present a *online memory re-templating technique* in Section 6.3 that offers fast correction of obsolete bit flip profile due to memory scrambling.

### 6.1 Multi-page Memory Massaging

In order to induce bit flips DNN model, memory massaging is required to map each victim page to a physical page whose vulnerable bit offset matches the offset of the targeted bit. In double-sided rowhammer, this includes a pre-step to set some of the attacker's pages in three consecutive rows in the same bank (*sandwich layout*), and the attacker should be aware of such memory layout. When the attacker's memory is properly situated, the vulnerable page positioning process starts.

---

[2]If one physical location has been chosen to flip model bit $i$, then it could not be utilized again for model bit $j$ even if both the page bit offset and the flip direction match.

**Massaging pre-step.** In order to get the *sandwich layout*, the attacker needs to be aware of both DRAM addressing and the physical addresses of its own pages. Based on our threat model, we assume that the adversary can not access privileged system interface including `/proc/pid/pagemap` for direct address translation. Our attacker can leverage previously proposed memory manipulating technique to force allocations of 2MB consecutive memory blocks [30]. Alternatively, the attacker can allocate a large chunk of memory in user-space, which will contain multiple sets of physically consecutive pages with a very high probability. We use row buffer side-channel as presented in [43] to reverse engineer the DRAM addressing function. The addressing function maps each page to one or multiple DRAM location pairs, denoted as *(row, set)*. The set number uniquely determines the *(channel, rank, bank)* combination for a specific physical address.

Once the attacker gains knowledge of its own physical page layout, the attacker reads the targeted chain of bits to flip. In our implementation, each targeted bit is represented as a three-element tuple ($vp_i$, $bop_i$, *mode*) where $vp_i$ is the targeted victim page, $bop_i$ is the targeted bit offset in that page. Finally *mode* indicates the desired flip direction and can be set to 0 (i.e., 1→0 flip) or 1 (i.e., 0→1 flip). In our attack instance where model weight file is the target, the page identifier is the serial number of the 4KB content that contains the targeted weight parameter. The attacker then checks all its own physical pages and looks for pages that have the targeted bit locations (i,e., *bop*). Flipping the targeted chain of bits is considered plausible with the attacker's current memory layout if each targeted page can be positioned and hammered independently. In case that certain vulnerable pages are not available, the attacker can verify the satisfiability for the next candidate set of bits.

#### 6.1.1 Compact Aggressors using in-row Pages

Conventionally, rowhammer attacks use full occupation of the two aggressor rows. However, preparing full aggressor rows for each target page unnecessarily wastes page utilization efficiency, and can also potentially increase the chance of failure for target page mapping. For instance, let's assume that one target page *pgid*1 needs to be positioned at *bank*0 and *row*10 while another target page *pgid*2 has be be placed at *bank*0 and *row*11. In this scenario, if we place *pgid*1 at *row*10, *row*9 and and *row*11 should be both locked as aggressor rows, making it impossible to map *pgid*2 to *row*11 at the same time. Since memory-exhaustion can raise alarm for potential rowhammer exploitation, it is critical for the attack to map target pages and also limit its memory footprint.

To improve page utilization and maximize chance of successful target page mapping, we develop a new rowhammering technique that creates compact aggressors. Our key observation is that rowhammering can manifest at finer-grained level: ***in-row pages***. An in-row page is the portion of a 4KB physical page that is mapped to one row. Figure 3 illustrates page to row mapping for two different memory configurations. As we can see, for a single channel single DIMM configuration, one physical page is mapped to one row, and thus each DRAM row contains two different physical pages. In a dual-channel memory setting, each page is split evenly to two in-row pages, and each DRAM row has four in-row pages (corresponding to four distinct physical pages).

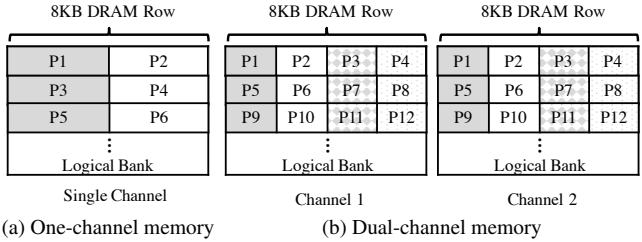

Figure 3: Physical page to row mapping on systems with two different memory configurations (**left**: single channel single DIMM/DDR3-Ivy Bridge; **right:** dual channel single DIMM/DDR3-Ivy Bridge).

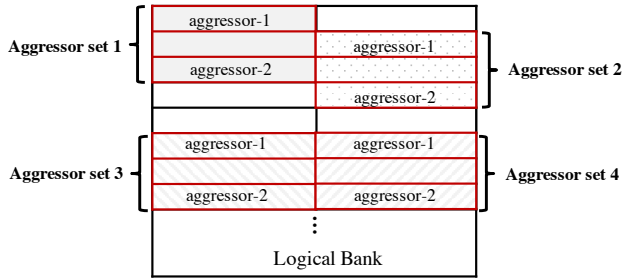

Figure 4: An example of attack memory preparation using compact aggressors. We illustrate four aggressor sets represented using different filled patterns.

The in-row page is the atomic hammering unit for each vulnerable page since other portions of the same page are mapped to different banks. As long as the in-row page right above and below the victim in-row page are setup and controlled as aggressors, the attacker is still able to induce the desired bit flip. More importantly, it has been observed that the cross-talk effect to a vulnerable memory cell is only attributed by *the bits in the adjacent rows at the same column*. In other words, different in-row pages can be hammered separately as long as the aggressor pages and victim pages are set properly. Our proposed attack leverages compact aggressor to prepare memory layout for rowhammering. Figure 4 illustrates a possible combination of aggressor settings considering a 4KB in-row page size (i.e., configuration in Figure 3a). We can observe that the victim page in aggressor set1 share the same DRAM row with the first aggressor in aggressor set2. Additionally, aggressor set3 and set4 occupies exactly the same consecutive rows, but they are able to induce bit flips without interference.

Obviously, this approach improves efficiency for page usage for the target page mapping phase.

### 6.1.2 Target Page Positioning

With the knowledge of compact aggressors, the attacker's next step is to find a mapping of each vulnerable page to the physical page in its memory space. We utilize a simple but effective heuristic algorithm that positions target pages with least number of satisfiable physical locations first. Once the mapping strategy is finalized, the attacker releases the corresponding physical pages and remaps the target page.

To accurately locate all the target pages, we take advantage of per-cpu page frame cache in Linux-based systems. Linux system uses the buddy system to manage page allocation. Memories are globally organized as zones by the buddy allocator. When a physical page is freed by a process running on certain CPU, the freed page is not immediately returned to the global memory pool. Instead, freed pages are pushed to a local fast page frame named `per-cpu pageset`. Later when the OS need to allocate a new page in the same hardware context, it will first attempt to get the page from the head of the list (i.e., stack-like access policy). Such design facilitates usage of pages that are still hot in private caches. Since the per-cpu page frame cache only manages pages locally, it has extremely low noise as compared to global memory pools. Note that When the number of pages frames in the list exceeds certain *recycling threshold*, a batch of pages are returned to the global pool maintained by buddy system.

We exploit per-cpu page frame cache to position the target pages in the following steps:

**Step 1:** The attacker determines the target page to exploitable physical page mapping for the targeted bit chains. Suppose we have $K$ bits to flip, we can denote the mapping as $(pgid_i, ppn_i)$, where $pgid_i$ represent the $i^{th}$ page in DNN's model weight memory and $ppn_i$ is the designated physical page frame for $pgid_i$. $i$ is within [1, $K$].

**Step 2:** The attacker frees the target physical pages $ppn_i$ from $ppn_1$ to $ppn_K$ in order using the `munmap` system interface. To avoid recycling of these pages to global pool, it is preferable that the number of pages freed ($K$) should be significantly less than the recycling threshold. In our testbed, we observe that the threshold is set to 180 by default, which is sufficient for our exploitation.

**Step 3:** Right after step 2, the attacker loads the target pages of the DNN model using `mmap`. The pages are loaded from $pgid_K$ to $pgid_1$. To avoid OS page pre-fetching that interrupt the page mapping, we use `fadvise` with the `FADV_RANDOM` after each `mmap` call. In the end, each target page is located to the attacker-controlled physical location.

## 6.2 Precise Rowhammering

Once the target pages are placed in the exploitable location, the attacker begins the initialization phase for the aggressor sets. Prior works typically uses the row-stripe patterns (i.e., 1-0-1 and 0-1-0) as they trigger most bit flips. However, on our platform we figured out that many physical pages exhibit multiple vulnerable locations (i.e., multiple bit flips). As mentioned in Section 5, the attacker needs to control the bit flips precisely at the targeted locations since extra bit flips undermine the effectiveness of our attack. Therefore, the attacker should avoid simultaneous bit flip at undesired page offsets. We design a precise rowhammering technique using a new data pattern called *column-page-stripe* data pattern. Using column-page-stripe data pattern, given that the victim row has bit sequence $b_0 b_1 ... b_j ... b_k ... b_n$ and assume that the goal is to flip bit $b_j$ and $b_k$, the attacker will set the content of the two aggressors to $b_0 b_1 ... \overline{b_j} ... \overline{b_k} ... b_n$. Particularly, we only configure the stripe pattern for the column where a bit flip is supposed to happen. For other bits that are expected to stay unchanged, the bits in its aggressors are kept the same as those in the victim page. Again, this strategy is built based on the fact that a bit flip is only controlled by bits in its aggressors that have the same column, and will not be influenced by the aggressor's bit values in other columns. With compact aggressors, the attacker configures the column-page-stripe pattern with the granularity of in-row page. Otherwise,

## 6.3 Online Memory Re-templating

Memory templating collects the profile of vulnerable bit location in DRAM modules. The validity of profile is based on the fact that a considerable amount of the bit flips are repeatable and stable. Our attack exploits those stable bit flips found in templating process. However, we observed that even for bit locations with stable flips, there are times (especially after system reboots) when our attack failed to toggle the value in the expected direction (e.g., 1→0). More interestingly, we found that such bit location almost always allows bit flip in the opposite direction (e.g., 0→1). Further investigation reveals that such phenomenon is related to memory scrambling. Memory scrambling is done by memory controller to encode data before they are sent to the DRAM modules [27]. The encoding scheme is based on a random seed that is set at boot time. Therefore, when system reboots, the memory controller may flip the logical representation of a bit to be stored in certain vulnerable cell. Accordingly, its bit flip orientation would change. While memory templating has been widely used before, we are not aware of any prior works that look at this issue. Note that the obsolescence of template is devastating for our proposed attack as it requires precise bit flips.

In order to address this problem, we augment the memory massaging process with an additional step. Specifically, before the attacker performs vulnerable page mapping (Sec-tion 6.1), it first quickly verifies whether its memory template has invalid flips for several stably vulnerable memory cells. This can be done by hammering a few pages in the attacker's own memory space. If expected bit flips are seen, the attacker knows that memory controller most likely has not changed its scrambling scheme yet, and thus the previous bit flip profile is still valid. Otherwise, the attack performs fast *online memory re-templating* that corrects the bit flip profile. It is worth noting that a complete templating of the attacker's memory space can take up to a few hours. Moreover, we figure out that complete profiling is not necessary. This is because no matter what data scrambling policy will be applied, the locations of the vulnerable memory cells would not change. Based on this observation, the attacker first filters out pages that do not have vulnerable bits at the desired locations (according to the targeted bit chain). This removes vast majority of pages in the re-templating process. For the rest of the pages, the attacker only needs to re-test its bit flip direction. Specifically, for each targeted page offset, the attacker exams its pages that have bit flips in that specific page offset regardless of whether 0→1 or 1→0 direction was recorded. The new direction is recorded and used to determine target page mapping.

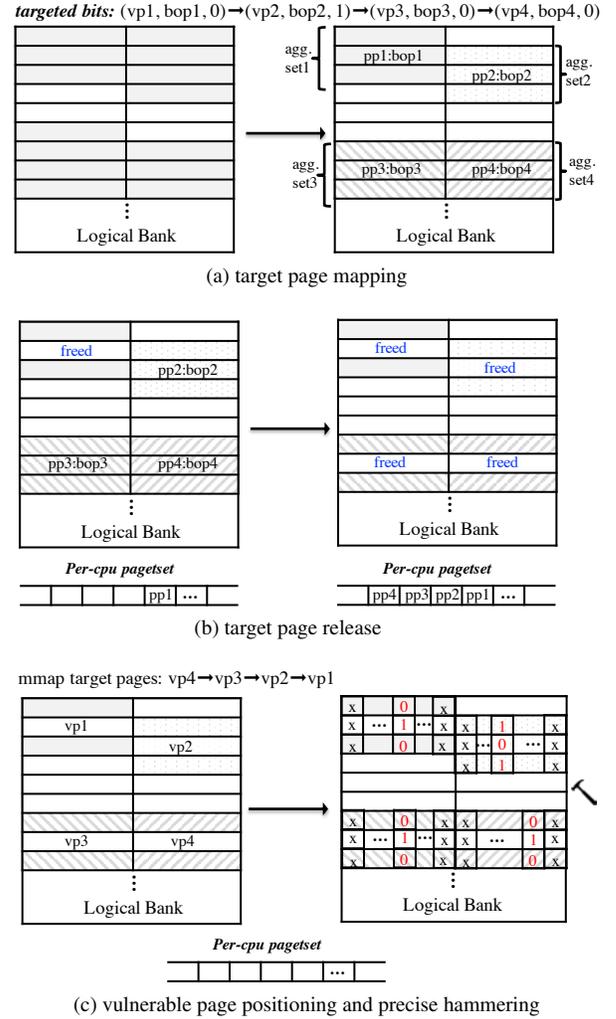

Figure 5: A step-by-step demo of DeepHammer attack.

### 6.4 Putting It All Together

By combining all the aforementioned rowhammer techniques, we build our DeepHammer framework. As a summary, we illustrate a step-by-step exploitation as shown in Figure 5. Figure 5a shows the process where the attacker prepares compact aggressor layout for all vulnerable pages. In this step, the attacker takes as input the targeted bits that are generated from our bit search algorithm as described in Section 5. The attacker is aware of the pages in its memory space that come with vulnerable bits at certain page offsets based on the bit profile. The attacker then prepares a mapping between the targeted pages to its physical pages, which will determine what page to release later. The attacker uses compact aggressors to improve memory usage efficiency. If the bit flip profile is obsolete due to memory scrambling, the attacker then performs a online memory re-templating process (not shown in this figure). Once vulnerable page to physical page mapping is determined and the compact aggressors are set, the attacker starts releasing the corresponding victim physical pages by exploiting the per-cpu page frame cache. In this illustration, the attacker releases the pages in the order: $pp1$, $pp2$, $pp3$, $pp4$ where $pp_i$ is the desired location to flip $bop_i$ in the target DNN's memory $vp_i$ (Figure 5b). After all target page frames are pushed to the per-cpu page frame cache, the attacker immediately loads the targeted victim pages in the reverse order as shown in Figure 5c: $vp_4$, $vp_3$, $vp_2$, $vp_1$. This achieves the expected mappings of $(vp_1, pp_1)$, $(vp_2, pp_2)$, $(vp_3, pp_3)$, $(vp_4, pp_4)$. Finally, the attack prepares the content of the aggressors to facilitate precise hammering using targeted column-page-stripe pattern. As shown in the right side of Figure 5c, to flip the bit at bit offset $bop_1$ from '0' to '1' in the target page $vp_1$, DeepHammer sets the stripe pattern $1-0-1$ only at one column that corresponds to $bop_1$. All the other columns in the aggressor set is set to $x-x-x$, which is a solid pattern that minimize inter-cell disturbances and avoid extra bit flip. When the aggressor are configured correctly, DeepHammer starts inducing bit flip at the four locations with doubled-sided rowhammering. In case that multiple aggressor sets are located in the same rows (maximum 2 for single channel and 4 for dual channel), DeepHammer can induce multiple targeted bit flips in one row hammering iteration. For example, since aggressor set3 and aggressor set4 share the same rows, DeepHammer only needs to hammer aggressor set3 because each DRAM access activates an entire row.

## 7 Experimental Setup

**Software setup.** Our deep learning platform was Pytorch 1.04 (python 3.6) which supports CUDA 9.0 for GPU acceleration. Our attack is evaluated with both computer vision and speech recognition applications. For object classification task in computer vision, several popular visual datasets, including Fashion-MNIST [61], CIFAR-10 [28] and ImageNet [11] are tested. Fashion-MNIST is the only gray-scale dataset we used, containing 10 classes of fashion dress images divided into 70k training images and 10k test images. CIFAR-10 contains 60K RGB images in size of $32 \times 32$. We follow the standard practice where 50K examples are used for training and the remaining 10K for testing. ImageNet dataset is a large dataset containing 1.2M training images divided into 1000 distinct classes. Images of size $224 \times 224$ are evenly distributed into 1000 output classes. For Fashion-MNIST, a simple LeNet [31] architecture was sufficient. For CIFAR-10 we evaluate on VGG-11, VGG-16 [50] ResNet-20 [17] and AlexNet [29]. For ImageNet classification we used ResNet-18, ResNet-34, ResNet-50 and memory frugal architecture including SqueezeNet [25] and MobileNet-V2 [47]. For speech recognition application, we test with a popular dataset, i.e. Google speech command [59] which is used for limited vocabulary speech recognition tasks. It has twelve output classes of voice command. We test the speech command dataset in VGG-11 and VGG-13 [50] architectures. For most of other analysis, we take the popular VGG-16 network [50] as the base benchmark. It is a 16-layer network among which 13 of them are convolutional networks and 3 fully connected layers. VGG-16 is a very dense network with close to 138 million parameters. The presence of parameters in heavy fully connected layers makes VGG-16 a highly dense network. The details of other architectures are included in the Appendix 11.

**Hardware setup.** Our DNN models are trained and analyzed on GeForce GTX 1080 Ti GPU platform. The GPU operates at a clock speed of 1481Mhz with 11GB dedicated memory. The trained model is deployed on a test-bed machine where our proposed attack is evaluated. The inference service runs on a Ivy Bridge-based Intel i7-3770 CPU that supports up to two memory channels. We have setup two different memory configurations for the machine. The first one is a single channel single DIMM setting with one 4GB DDR3 memory as shown in Figure 3a, and the second configuration features a dual-channel single DIMM setting with two 4GB DDR3 memory modules.

**Memory templating.** We used the DRAM addressing scheme as released in [43]. Using the addressing function, the attacker perform memory templating by scanning the rows in the target DRAM modules. Each bank in the DRAM has 32768 rows, and each DRAM DIMM has 16 banks. We observe that bit flips are uniformly distributed across banks. Our attack randomly samples rows in each of the bank. It is worth noting that while templating is an offline process, it is important that it does not corrupt the system to avoid raising security alarms. Therefore, the attacker skips rows that are close to physical pages not belonging to itself.

| Dataset | Architecture | Network Parameters | Acc. before Attack (%) | Random Guess Acc. (%) | Acc. after Attack (%) | Min. # of Ait-flips |
|---|---|---|---|---|---|---|
| Fashion MNIST | LeNet | 0.65M | 90.20 | 10.00 | 10.00 | 3 |
| Google Speech Command | VGG-11 | 132M | 96.36 | 8.33 | 3.43 | 5 |
| | VGG-13 | 133M | 96.38 | | 3.25 | 7 |
| CIFAR-10 | ResNet-20 | 0.27M | 90.70 | 10.00 | 10.92 | 21 |
| | AlexNet | 61M | 84.40 | | 10.46 | 5 |
| | VGG-11 | 132M | 89.40 | | 10.27 | 3 |
| | VGG-16 | 138M | 93.24 | | 10.82 | 13 |
| ImageNet | SqueezeNet | 1.2M | 57.00 | 0.10 | 0.16 | 18 |
| | MobileNet-V2 | 2.1M | 72.01 | | 0.19 | 2 |
| | ResNet-18 | 11M | 69.52 | | 0.19 | 24 |
| | ResNet-34 | 21M | 72.78 | | 0.18 | 23 |
| | ResNet-50 | 23M | 75.562 | | 0.17 | 23 |

Table 1: Results of vulnerable bit search on different applications, datasets and DNN architectures.

# 8 Evaluation

In this section, we present the evaluation results to show the effectiveness of our proposed DeepHammer attack.

**Bit flip profile.** To extract most of the bit flips from the target DRAM module, doubled-sided rowhammering with row-stripe data pattern (1-0-1 and 0-1-0) are utilized. We first perform an exhaustive test by hammering rows in all the banks. We configure the hammering time for each row to be 190ms, which is sufficiently long to induce bit flips in vulnerable cells. In the memory template phase, we observe 2.2 bit flips every second. Overall, we found that each bank contains $35K$ to $47K$ bit flips. Templating each bank takes about 5 hours. We further observe that more than 60% of the vulnerable physical pages have at least two flippable memory cells. This highlights the need to perform precising rowhammering using our proposed targeted column-page-strip pattern.

Based on our experiment, it takes about 120 seconds for our flip-aware bit searching algorithm to generate one candidate. Note that since bit search can be done offline, it is not time-critical as compared to the online exploitation phase. The attacker's objective is to completely malfunction a well-trained DNN model, i.e. degrading its inference accuracy to the level of random guess. Therefore, the ideal attack successful accuracy will be close to (1/# of output classes)×100%. Apparently, the target accuracy after attack would be different for distinct datasets. For instance, CIFAR-10 and ImageNet have 10 and 1000 output classes, thus the expected inference accuracy after attacks would be around 10% and 0.1%, respectively. Table 1 demonstrates the identified bit flips and attack results if all bits are flipped among 12 different architecture/dataset configurations. As shown in the figure, DeepHammer is able to successful compromise all the networks using maximum 24 bit flips. We also observe that the required number of bit flips fluctuates significantly across different architecture/dataset combinations. We note that the vulnerability to model bit flips can potentially be affected by both network size and network topology. Specifically, for the CIFAR-10 dataset, with a larger network size, VGG-16 has demonstrated relatively higher robustness as compared to VGG-11 (13 required bit flips vs. 3 bit flips). Such observation aligns with previous studies on adversarial input attack [40] showing potential improvement of model robustness with increasing network size. Additionally, from network topology perspective, the ResNet architecture family has consistently demonstrated better resilience to model bit flips with more than 20 required bits to successfully compromise these models. We hypothesize that such characteristic may be due to the existence of the residual connection in the networks (See Section 9.2). For compact networks, we observe that MobileNet-V2 is extremely vulnerable on the ImageNet dataset where DeepHammer only need to trigger 2 bit flips to succeed, which is considerably less than the other compact network, SqueezeNet. Note that MobileNet-V2 has several distinguishing aspects in terms of network topology and size: 1) The MobileNet architecture family is different from the other DNN models with the presence of the combined depth-wise separable convolution and point-wise convolution layer; 2) It has deep network architecture with 54 layers while hosting relatively small amount of model parameters. We envision that network size and topology have an interplay in terms of influencing the vulnerability of DNN models. Finally, besides computer vision application, DeepHammer is also capable of compromising VGG-11 and VGG-13 on Google speech command dataset, which reveals that our proposed attack is effective for a wide range of DNN models and application domains.

Note that, our searching algorithm could generate multiple bit chains to attack one network. We reported the minimum number of bits required in Table 1. Table 2 illustrates three identified bit chains from our searching algorithm to attack VGG-16 in CIFAR10 dataset. Due to space limit, more iden-tified bit chain samples for other network architectures are shown in Table 4 of Appendix D. We observe that, to completely compromise the target quantized VGG-16, DeepHammer only needs to attack as few as *13* bits. We also observe that in terms of bit flip direction (i.e., *mode*), more than 70% of the vulnerable bits use 1→0 flip. Such high disparity is because, in a typical DNN model, vast majority of the weights are 0s while the non-zero weights play a key role in determining the classification output. Therefore, to maximize accuracy drop, modifying non-zero weights at proper locations can considerably change the prediction behavior.

| # of Bits | Identified chain of bit flips (page#, bop, mode) | Hammer time (s) | Accuracy (%) |
|---|---|---|---|
| 13 | **c1:** (1,4847,0)→(8,25719,0)→(4,23111,0)→(20,7887,0)→(128,3047,0)→(10,1623,0)→(13,2247,0) →(2,16447,1)→(9,22079,1)→(356,16823,0)→(60,11655,0)→(3,2087,1)→(3720,29048,0) | **66** | 10.82 |
| 18 | **c2:** (1,11335,0)→(8,223,0)→(28,12567,0)→(7,743,1)→(2,17127,0)→(10,3135,1)→(91,9527,0)→ (24,28447,1)→(9,13535,1)→(6,30071,1)→(3720,28728,0)→(15,28431,1)→(460,24375,0)→(154,20671,0) →(92,32103,0)→(48,12767,1)→(157,15023,0)→(16,27911,1) | **82** | 10.70 |
| 20 | **c3:** (9,12839,0)→(1,9367,0)→(17,9687,0)→(4,20031,0)→(70,17479,0), (25,975,0), (229,9199,0)→ (24,31287,0)→(14,11247,0)→(183,5167,0)→(55,12063,0)→(62,9111,0)→(29,25391,0)→(3720,16248,1) →(2792,1192,0)→(395,30063,0)→(706,4200,1)→(292,19583,0)→(28,21263,0)→(431,20550,1) | **96** | 10.88 |

Table 2: List of three candidate bit chains (i.e., c1, c2 and c3) to attack VGG16 generated by our flip-aware bit search algorithm.

Another critical observation is that the targeted weight bits mostly cluster in *the first and last a few layers*. For instance, for VGG-16, half of the 13 targeted bit flips (Table 2) are located in the front-end of the network. Additionally, all the 3 bit flips in VGG-11 network are located in the last 3 layers. This potentially indicates that the first and last layers of DNN models are more vulnerable to model weight bit flips. Based on prior studies and our findings, we believe this is because perturbations in early stages of DNN can get propagated and thus amplified significantly towards the end, while on the other hand, changes of model parameters at the back-end of the network can directly alter the classification outcome.

**DeepHammer online exploitation** The online exploitation phase is implemented as a standalone process. We run DeepHammer to target each of the three bit chains as demonstrated in Table 2. We observe that in the target page mapping process, the exploitable pages found for each bit in the chain are almost the same, indicating that the offset filtering scheme in our bit search algorithm is highly effective. In order to find aggressor sets for all the targeted bits, DeepHammer needs to allocate a relatively large chunk of memory. Our experiments show that to satisfy target page mapping for multiple targeted pages, DeepHammer has to allocate around 12% of the system memory. Apparently, the size of allocation depends on number of desirable bits to flip. Our profiling test shows that allocation of 20% system memory almost always guarantee satisfaction of mapping. We note that such memory allocation can succeed most of the time in the system without triggering out-of-memory exceptions (unless the available system memory is extremely low). Additionally, our attack only holds the memory for target page mapping (the step shown in Figure 5c). After the mapping is completed, the attacker can then release the vast majority of memory pages that are not needed anymore, making it unlikely for system underlying security policy to raise alarms.

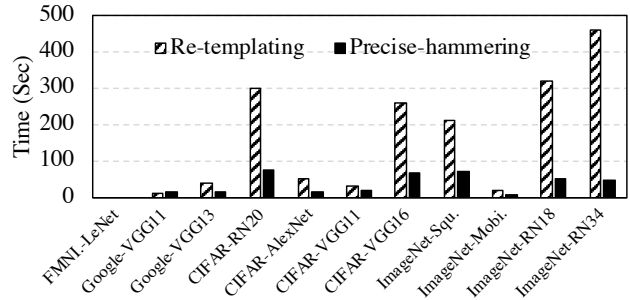

Figure 6: DeepHammer re-templating time and multi-bit hammering time for all dataset/architecture combinations.

Table 2 also presents the online exploitation performance for VGG-16 under the three candidate bit chains. For all the three runs, our proposed attack is able to achieve the goal of degrading the inference accuracy of the target DNN to about 10%. Due to variations in test dataset, the actual achieved accuracy is slightly higher (e.g., 10.82% for *c*2). We observe that DeepHammer can perform target page mapping and precise rowhammering very fast. All three attack instances requires less than 100 seconds to induce bit flips. The high attack efficiency is because DeepHammer uses per-cpu page frame buffer that allows fast remapping of previously released pages in a deterministic way. This avoids the process of page relocation that can take substantially longer. We also demonstrate the time it takes for memory-retemplating in case the bit flip profile is obsolete. We can see that while memory-templating can take hours to finish, our online memory-retemplating can finish in a few minutes. This is because we only need to check the pages with vulnerable memory cells at the desirable locations. Since memory scrambling only changes the flip direction, but not the vulnerable bit locations, our attack effectively leverages the vulnerable bit locations in the old bit profile to filter out most of pages without templating on them. Additionally, Figure 6 illustrates the online rowhammering time as well as memory re-templating times for all model and dataset combinations. We observe that all attack instances can succeed within less than 10 minutes. Overall, even considering memory re-templating, DeepHammer is still able to compromise the target quantized DNN in minutes for the online phase.

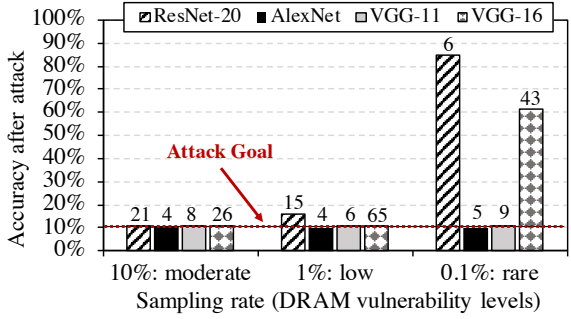

Figure 7: Attack results for DRAMs with different vulnerability levels via sampling. The number on top of each bar denotes the number of bit flips needed for each configuration.

**Impact of DRAM vulnerability.** Our memory templating phase has identified about 600K bit flips in DRAM module, which shows that the underlying DRAM modules are highly vulnerable to rowhammer exploitation. To understand the impact of DRAM vulnerability on the effectiveness of DeepHammer, we further perform a sensitivity study. Specifically, as vulnerable bit locations are often uniformly distributed in DRAMs [13], we randomly sample the bit profile at three different rates (10%: *moderate* amount of flips, 1% *low* amount of flips, 0.1%: *rare* flips), which matches a wide spectrum of realistic DRAM vulnerability levels according to prior study in [54] [3]. Note that DeepHammer is designed to work effectively with *partial knowledge* of bit flip patterns (e.g., number of vulnerable bits in one page) due to the use of the precise hammering technique that ensures only bits at the locations the attacker is aware of would be flipped. Figure 7 demonstrates the attack results on 4 different models using CIFAR-10 dataset. We can see that the attack on AlexNet, VGG-11 and VGG-16 are successful almost for all cases except that the desired accuracy is not reached for VGG-16 at 0.1% sampling rate. Additionally, for ResNet-20, the achieved accuracy is slightly higher than attack goal under 1% sampling (15.75% vs. 10.00%), and the attack is not successful under 0.1% sampling. We note that this is because ResNet-20 has *smallest network size* (See Table 2) with very low number of physical pages to exploit. Therefore, under less vulnerable DRAM configurations, the number of bits that can be practically flipped is heavily constrained by the bit profile in the target system, making it hard for our search algorithm to target top-ranked model bits. Differently, our investigation shows that if the system constraint is not modeled, a theoretical attack can succeed using 20 bit flips in ResNet-20. We note that this highlights the importance of our proposed flip-aware bit search scheme with respect to understanding empirical danger of bit fault attacks against DNN in real system.

---

[3] We choose to sample our existing bit profile instead of directly using existing bit flip database in [54] so that we can empirically demonstrate the result of the attack.

**Impact of single-sided hammering.** We also studied the effect of using single-sided rowhammer that does not require locating of two aggressor rows. On the same machine, we figured out that with single-sided rowhammering, much less vulnerable bits are found (**1876** 0→1 flips and **1468** 1→0 flips). We tested the same 4 models used for the aforementioned DRAM vulnerability study. We find that results of DeepHammer using singled-sided hammering is similar to doubled-sided rowhammering under the lowest vulnerability level. Specifically, our attack on AlexNet and VGG-11 succeeded with 7 and 6 bit flips, respectively while the desired accuracy drop is not achieved for VGG-16 and ResNet-20. Such results are expected since the number of total exploitable bits are about 0.3% compared to doubled-sided rowhammering.

## 9 Discussion

### 9.1 Untargeted and Targeted Attacks

DeepHammer mainly focuses on untargeted attacks that degrades the overall inference accuracy to the close-to-random-guess level without explicitly controlling the specific output class. However, we do have some useful observations that could lead to a potential targeted attack. *Observation-1:* The identified bit-flip chain forces almost all the inputs to be classified into one particular output group, instead of completely random, even though the test batch chosen to calculate gradient is random and may contain inputs from different groups. We call this particular output as *winner-group*. *Observation-2:* We did not intentionally choose the winner-group in our original method, thus DeepHammer does not control the winner-group directly. However, we find that the winner-group is heavily dependent on which group of input sample batch is used to compute the bit gradients. This is likely because our search algorithm mainly follows the gradient-descend direction to amplify particular weights that are strongly linked to one particular output group. Thus, the test data in different groups may help us find different weights strongly connected to the corresponding output groups, which could enable controlling of the winner-group by the adversary. These observations motivate us to find a way of *extending our attack to a variant of targeted attack*: forcing DNN to classify any input to one target group if the attacker can provide one batch of test data belonging to the target group to our search algorithm.

To validate this targeted attack extension, we test ResNet-20 on CIFAR 10 dataset. To target class-1, we intentionally choose a test batch with all images from class-1 to perform our flip-aware bit search. It shows that almost 99.63% of all test inputs will be classified into class-1 with just 18 bit-flips. Similar results are observed in all other groups (e.g., class-9 targeted attack requires 19 bit-flips). We will investigate further in our future work about other types of targeted attacks, e.g. only misclassifying certain inputs to specific classes without influencing the rest of inputs.

| Architecture: | Ori. Acc. (%) | After Attack Acc. (%) | # of Bit-Flips |
|---|---|---|---|
| ResNet-20 | 90.7 | 10.92 | 21 |
| ResNet-20×2 | 92.01 | 14.19 | 30 |

Table 3: Ablation study of model redundancy.

## 9.2 Potential Mitigation Techniques

**DNN algorithm level mitigation.** Prior works have shown that wide DNNs are typically more robust to noise injection for adversarial inputs [18, 40]. As DeepHammer can be considered as one kind of attack that injects noises to network weights, we expect wider networks could be more resilient to such attack. To validate this hypothesis, we evaluate the effectiveness of DeepHammer for both standard ResNet-20 and double width ReseNet-20×2. From Table 3, we can see that DeepHammer requires higher number of flips as we increase its network width by 2×. In contrast to ResNet-20 baseline model which requires only 21 flips to reach 10.92 % accuracy, the ResNet-20 (×2) model accuracy degrades to 14.19% even after 30 flips. Apparently, increasing the network width (i.e. redundant model) alleviates the effect of DeepHammer (at the cost of increasing number of network parameters), although it does not completely thwart DeepHammer. Furthermore, based on results of different network architectures shown in Table 1, we find that ResNet family with residual connections are in general more robust, requiring more number of bit flips, but still cannot completely defend against DeepHammer. In complete contrast to other deeper networks which come at the expense of gradient vanishing [20], ResNet's residual connections make the network's learning process more robust and resilient.

**Protecting top-N vulnerable bits in model.** One straightforward solution is to identify $n$ most vulnerable bits and selectively protect these bits by system software. For example, in Round-$i$ (Ri), we can apply the proposed GBR algorithm to identify vulnerable n-bits that degrades the DNN accuracy close to random guess (10% for CIFAR-10), then those vulnerable bits are assumed to be protected by OS and labeled as bits that cannot be flipped in round-$(i + 1)$. Note that we do not consider the memory layout and flip-aware bit search to perform this analysis. As shown in Figure 8, we run the experiments ten rounds, where it does not show significant

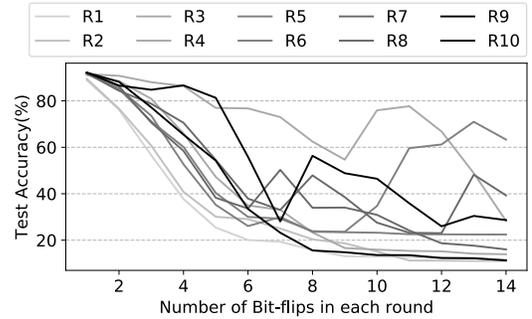

Figure 8: Test accuracy versus Number of bit-flips of VGG-16 on CIFAR-10. Curve in darker color indicates later round.

performance degradation of the attack when more bits are secured.

The above observation clearly shows that the search space of vulnerable bits is too large. Thus protecting only small amount of those vulnerable bits may not be a feasible solution. As a result, defense solutions which provides both software- and hardware-level guarantee of data integrity may be one possible direction for future investigation.

**Hardware-based protections against model tampering.** Another direction is to leverage hardware support to avoid data tampering on vulnerable/untrusted memory modules. Several recent works have studied the use of secure enclave (e.g., Intel SGX [10]) to protect the privacy of critical data in DNN such as sensitive user inputs, training data and model parameters [24, 56]. SGX-based solution also offers data integrity protection against off-chip data tampering. While such approaches can work on small DNN models, Intel SGX-based techniques are subject to high performance overhead for maintaining large models in enclaves [16]. This could pose serious impact on ML applications that are latency critical. On the other hand, while many vulnerable bits exist in DNN model, our investigation has revealed that the identified bits are mostly concentrated in the *first* and *last* a few layers (See Appendix D). Therefore, securing these vulnerable layers instead of the entire model may efficiently improve the robustness of DNN models with low overhead. Particularly, one promising solution is to selectively preload these critical model layers onto CPU caches. In this way, even some bits are corrupted in the DRAM, it will not adversely influence the inference accuracy of the target model. We note that there are already commercial-off-the-shelf supports that enable allocation of dedicated cache regions to applications for Quality-of-Service purposes (e.g, Intel CAT [26]). System administrator can take advantage of this feature to lock vulnerable model layers to prevent from tampering while not incurring considerable runtime overhead.

## 9.3 Limitations and Future Work

We note that in our current implementation, we only exploit a single bit flip in each physical page for our proposed attack. However, it is possible to leverage DRAM pages that have multiple bit locations in one page. Note that the use of multi-bit flip in a physical page is beneficial in terms of attack efficiency. This is because we find that the highly influential bits in DNN models are not distributed uniformly. Particularly, the first and last a few layers have more vulnerable bits where many of them are located in the same 4KB page boundary. Therefore, multiple-bit flip in the same page can reduce the number of targeted pages as well as the required row hammering iterations. We plan to explore such optimization in future work.

Our threat model assumptions are similar to conventional white-box attack approach in related domain [18, 40]. Under such assumption, an adversary has access to the network architecture, weight values and one batch of test data. While such information can be potentially gained as discussed in Section 3, such requirement may not be applicable in all scenarios. Nevertheless, to address such limitation, in our future work, we will explore ways to perform our attack in a semi-black box setup without knowing the precise weights of a victim DNN model. Note that architecture information is relatively easier to obtain due to the fact that many applications directly adapt popular network architectures to do fine-tuning. Additionally, the adversary could potentially perform this semi-black box attack by training a substitute model through label querying of the target model and transferring the attack from the substitute model to the target model.

## 10 Related Work

Machine learning have been increasingly adopted in a variety of application domains [8, 19, 32–35, 49, 58, 60]. Deep neural work is one of the most promising ML techniques due to its superior performance. Several recent works have focused on the vulnerability of DNN models [18, 36, 40]. Traditional security of DNN mainly focused on the adversarial input attack where an adversary maliciously perturbed an input image with an intention to degrade the performance of the DNN [40, 52]. Some of the recent works have shifted the paradigm towards neural network's parameters integrity. Inserting hardware trojan and causing networks to behave maliciously has been effectively demonstrated by few recent works [37, 38] as well. However, attacking neural network weight parameters has received far less attention.

But recently some of the fault injection attack methods have demonstrated serious vulnerability in DNN models [5, 36, 44]. Fault injection methods have proved to be effective in completely destroying the functionality of the DNN models. [36] proposed a single bias attack which changes the value of a certain bias to an extremely large value to cause the DNN malfunction. Others have proposed activation level fault injection [3]. Attacking the DNN weight parameters of a full precision model through practical row hammer attack is proposed by [21]. However, attacking a full-precision DNN is relatively easy than attacking a fixed precision model as shown by [21, 44]. Thus our proposed DeepHammer work is the first end to end system level practical attack of fixed point quantized weights on DNN model.

Several works have been proposed to mitigate rowhammer attacks. These defense mechanisms attempt to capture/stop one or multiple necessary steps taken in a rowhammer attack. Specifically, to avoid fast access to DRAM, some systems can intentionally disable *clflush* instructions that allow memory requests to bypass caches [57]. To prevent memory row proximity to critical data structure such as kernel space memory, OS supports are proposed to isolate user-space DRAM rows from kernel DRAM rows through DRAM partitioning [4]. Additionally, many existing rowhammer attacks use memory spraying in order to force victim pages to vulnerable DRAM locations, this leads to memory exhaustion that can be detected by system security policies [39]. Hardware-based protection mechanisms such as Targeted Row Refresh (TRR) monitors DRAM row access and refreshes a DRAM row that is potentially under attack [41]. ECC memories can potentially detect and correct rowhammer-induced bit flips. However, recent works have demonstrated that bit flips are still possible even with the presence of such hardware-based protections [9, 57].

## 11 Conclusion

In this paper, we present a new class of hardware-based attack called DeepHammer to deplete the intelligence of quantized deep neural networks. Our key observation is that while single-bit flip attack is effective in degrading the accuracy for full-precision DNN models, quantized deep neural networks are inherently robust to such attacks. DeepHammer leverages the rowhamer vulnerability to significantly degrade the inference accuracy of the target model by deterministically inducing a chain of bit flips in DNN model weights. We implement a novel flip-aware bit search technique that is able to locate the least amount of vulnerable bits in the target DNN models for flipping. We further design several new rowhammer techniques to enable fast and precise flipping the the targeted chain of bits. We evaluate DeepHammer on real system setup and provide an end-to-end demonstration of the attack. Our evaluation results show that DeepHammer is able to compromise the target DNN models within minutes. Our work highlight the need to develop tamper-resistant deep neural models to tackle future hardware-based fault injection attacks.

# References

[1] Model zoo: Discover open source deep learning code and pretrained models, 2019. https://modelzoo.co.

[2] Lejla Batina, Shivam Bhasin, Dirmanto Jap, and Stjepan Picek. CSI neural network: Using side-channels to recover your artificial neural network information. *CoRR*, abs/1810.09076, 2018.

[3] Battista Biggio, Luca Didaci, Giorgio Fumera, and Fabio Roli. Poisoning attacks to compromise face templates. In *International Conference on Biometrics*, pages 1–7. IEEE, 2013.

[4] Ferdinand Brasser, Lucas Davi, David Gens, Christopher Liebchen, and Ahmad-Reza Sadeghi. Can't touch this: Software-only mitigation against rowhammer attacks targeting kernel memory. In *USENIX Security Symposium*, pages 117–130, 2017.

[5] J Breier, X Hou, D Jap, L Ma, S Bhasin, and Y Liu. Deeplaser: Practical fault attack on deep neural networks. *ArXiv e-prints*, 2018.

[6] Wayne Burleson, Onur Mutlu, and Mohit Tiwari. Who is the major threat to tomorrow's security? you, the hardware designer. In *IEEE Design Automation Conference*, pages 1–5. IEEE, 2016.

[7] Nicholas Carlini and David Wagner. Towards evaluating the robustness of neural networks. In *IEEE Symposium on Security and Privacy*, pages 39–57. IEEE, 2017.

[8] Chenyi Chen, Ari Seff, Alain Kornhauser, and Jianxiong Xiao. Deepdriving: Learning affordance for direct perception in autonomous driving. In *IEEE International Conference on Computer Vision (ICCV)*, pages 2722–2730. IEEE, 2015.

[9] Lucian Cojocar, Kaveh Razavi, Cristiano Giuffrida, and Herbert Bos. Exploiting correcting codes: On the effectiveness of ecc memory against rowhammer attacks. *IEEE Symposium on Security and Privacy*, 2019.

[10] Victor Costan and Srinivas Devadas. Intel sgx explained. *IACR Cryptology ePrint Archive*, 2016(086):1–118, 2016.

[11] Jia Deng, Wei Dong, Richard Socher, Li-Jia Li, Kai Li, and Li Fei-Fei. Imagenet: A large-scale hierarchical image database. In *IEEE Conference on Computer Vision and Pattern Recognition*, pages 248–255. IEEE, 2009.

[12] Hongyu Fang, Sai Santosh Dayapule, Fan Yao, Miloš Doroslovački, and Guru Venkataramani. Prefetch-guard: Leveraging hardware prefetches to defend against cache timing channels. In *IEEE International Symposium on Hardware Oriented Security and Trust*, pages 187–190. IEEE, 2018.

[13] Daniel Gruss, Moritz Lipp, Michael Schwarz, Daniel Genkin, Jonas Juffinger, Sioli O'Connell, Wolfgang Schoechl, and Yuval Yarom. Another flip in the wall of rowhammer defenses. In *IEEE Symposium on Security and Privacy*, pages 245–261. IEEE, 2018.

[14] Seungyeop Han, Haichen Shen, Matthai Philipose, Sharad Agarwal, Alec Wolman, and Arvind Krishnamurthy. Mcdnn: An approximation-based execution framework for deep stream processing under resource constraints. In *International Conference on Mobile Systems, Applications, and Services*, pages 123–136. ACM, 2016.

[15] Song Han, Huizi Mao, and William J Dally. Deep compression: Compressing deep neural networks with pruning, trained quantization and huffman coding. *arXiv preprint arXiv:1510.00149*, 2015.

[16] Lucjan Hanzlik, Yang Zhang, Kathrin Grosse, Ahmed Salem, Max Augustin, Michael Backes, and Mario Fritz. Mlcapsule: Guarded offline deployment of machine learning as a service. *arXiv preprint arXiv:1808.00590*, 2018.

[17] Kaiming He, Xiangyu Zhang, Shaoqing Ren, and Jian Sun. Deep residual learning for image recognition. In *IEEE Conference on Computer Vision and Pattern Recognition*, pages 770–778. IEEE, 2016.

[18] Zhezhi He, Adnan Siraj Rakin, and Deliang Fan. Parametric noise injection: Trainable randomness to improve deep neural network robustness against adversarial attack. In *IEEE Conference on Computer Vision and Pattern Recognition*. IEEE, 2019.

[19] Geoffrey Hinton, Li Deng, Dong Yu, George E Dahl, Abdel-rahman Mohamed, Navdeep Jaitly, Andrew Senior, Vincent Vanhoucke, Patrick Nguyen, and Tara N Sainath. Deep neural networks for acoustic modeling in speech recognition: The shared views of four research groups. *IEEE Signal Processing Magazine*, 29(6):82–97, 2012.

[20] Sepp Hochreiter. The vanishing gradient problem during learning recurrent neural nets and problem solutions. *International Journal of Uncertainty, Fuzziness and Knowledge-Based Systems*, 6(02):107–116, 1998.

[21] Sanghyun Hong, Pietro Frigo, Yigitcan Kaya, Cristiano Giuffrida, and Tudor Dumitras. Terminal brain damage: Exposing the graceless degradation in deep neural networks under hardware fault attacks. *CoRR*, abs/1906.01017, 2019.

[22] Itay Hubara, Matthieu Courbariaux, Daniel Soudry, Ran El-Yaniv, and Yoshua Bengio. Binarized neural networks. In *Advances in neural information processing systems*, pages 4107–4115, 2016.

[23] Itay Hubara, Matthieu Courbariaux, Daniel Soudry, Ran El-Yaniv, and Yoshua Bengio. Quantized neural networks: Training neural networks with low precision weights and activations. *The Journal of Machine Learning Research*, 18(1):6869–6898, 2017.

[24] Tyler Hunt, Congzheng Song, Reza Shokri, Vitaly Shmatikov, and Emmett Witchel. Chiron: Privacy-preserving machine learning as a service. *arXiv preprint arXiv:1803.05961*, 2018.

[25] Forrest N Iandola, Song Han, Matthew W Moskewicz, Khalid Ashraf, William J Dally, and Kurt Keutzer. Squeezenet: Alexnet-level accuracy with 50x fewer parameters and< 0.5 mb model size. *arXiv preprint arXiv:1602.07360*, 2016.

[26] Intel. Introduction to Cache Allocation Technology in the Intel® Xeon® Processor E5 v4 Family, 2016. https://software.intel.com/en-us/articles/introduction-to-cache-allocation-technology.

[27] Yoongu Kim, Ross Daly, Jeremie Kim, Chris Fallin, Ji Hye Lee, Donghyuk Lee, Chris Wilkerson, Konrad Lai, and Onur Mutlu. Flipping bits in memory without accessing them: An experimental study of dram disturbance errors. In *International Symposium on Computer Architecture*, pages 361–372. IEEE Press, 2014.

[28] Alex Krizhevsky, Vinod Nair, and Geoffrey Hinton. Cifar-10 (canadian institute for advanced research). http://www.cs.toronto.edu/kriz/cifar.html, 2010.

[29] Alex Krizhevsky, Ilya Sutskever, and Geoffrey E Hinton. Imagenet classification with deep convolutional neural networks. In *Advances in Neural Information Processing Systems*, pages 1097–1105, 2012.

[30] Andrew Kwong, Daniel Genkin, Daniel Gruss, and Yuval Yarom. Rambleed: Reading bits in memory without accessing them. In *IEEE Symposium on Security and Privacy*. IEEE.

[31] Yann LeCun et al. Lenet-5, convolutional neural networks. *URL: http://yann. lecun. com/exdb/lenet*, 20:5, 2015.

[32] Li Li, Miloš Doroslovački, and Murray H. Loew. Discriminant analysis deep neural networks. In *53rd Annual Conference on Information Sciences and Systems*, pages 1–6, March 2019.

[33] Li Li, Miloš Doroslovački, and Murray H. Loew. Loss functions forcing cluster separations for multi-class classification using deep neural networks. In *IEEE ASILOMAR Conference*, pages 1–5. IEEE, Nov 2019.

[34] Chong Liu and Hermann J Helgert. Beammap: Beamforming-based machine learning for positioning in massive mimo systems.

[35] Chong Liu and Hermann J Helgert. An improved adaptive beamforming-based machine learning method for positioning in massive mimo systems.

[36] Yannan Liu, Lingxiao Wei, Bo Luo, and Qiang Xu. Fault injection attack on deep neural network. In *IEEE/ACM International Conference on Computer-Aided Design*, pages 131–138. IEEE, 2017.

[37] Yingqi Liu, Shiqing Ma, Yousra Aafer, Wen-Chuan Lee, Juan Zhai, Weihang Wang, and Xiangyu Zhang. Trojaning attack on neural networks. 2017.

[38] Yingqi Liu, Shiqing Ma, Yousra Aafer, Wen-Chuan Lee, Juan Zhai, Weihang Wang, and Xiangyu Zhang. Trojaning attack on neural networks. In *25nd Annual Network and Distributed System Security Symposium*, 2018.

[39] Xiaoxuan Lou, Fan Zhang, Zheng Leong Chua, Zhenkai Liang, Yueqiang Cheng, and Yajin Zhou. Understanding rowhammer attacks through the lens of a unified reference framework. *arXiv preprint arXiv:1901.03538*, 2019.

[40] Aleksander Madry, Aleksandar Makelov, Ludwig Schmidt, Dimitris Tsipras, and Adrian Vladu. Towards deep learning models resistant to adversarial attacks. In *International Conference on Learning Representations*, 2018.

[41] Janani Mukundan, Hillery Hunter, Kyu-hyoun Kim, Jeffrey Stuecheli, and José F Martínez. Understanding and mitigating refresh overheads in high-density ddr4 dram systems. In *ACM SIGARCH Computer Architecture News*, volume 41, pages 48–59. ACM, 2013.

[42] Nina Narodytska and Shiva Prasad Kasiviswanathan. Simple black-box adversarial perturbations for deep networks. *arXiv preprint arXiv:1612.06299*, 2016.

[43] Peter Pessl, Daniel Gruss, Clémentine Maurice, Michael Schwarz, and Stefan Mangard. DRAMA: Exploiting DRAM addressing for cross-cpu attacks. In *USENIX Security Symposium*, pages 565–581, 2016.

[44] Adnan Siraj Rakin, Zhezhi He, and Deliang Fan. Bit-flip attack: Crushing neural network with progressive bit search. In *The IEEE International Conference on Computer Vision*, October 2019.

[45] Kaveh Razavi, Ben Gras, Erik Bosman, Bart Preneel, Cristiano Giuffrida, and Herbert Bos. Flip feng shui: Hammering a needle in the software stack. In *USENIX Security Symposium*, pages 1–18, Austin, TX, August 2016. USENIX Association.

[46] Mauro Ribeiro, Katarina Grolinger, and Miriam AM Capretz. Mlaas: Machine learning as a service. In *IEEE International Conference on Machine Learning and Applications*, pages 896–902. IEEE, 2015.

[47] Mark Sandler, Andrew Howard, Menglong Zhu, Andrey Zhmoginov, and Liang-Chieh Chen. Mobilenetv2: Inverted residuals and linear bottlenecks. In *IEEE Conference on Computer Vision and Pattern Recognition*, pages 4510–4520, 2018.

[48] Mark Seaborn and Thomas Dullien. Exploiting the dram rowhammer bug to gain kernel privileges. *Black Hat*, 15, 2015.

[49] B. Shickel, P. J. Tighe, A. Bihorac, and P. Rashidi. Deep ehr: A survey of recent advances in deep learning techniques for electronic health record (ehr) analysis. *IEEE Journal of Biomedical and Health Informatics*, 22(5):1589–1604, Sep. 2018.

[50] Karen Simonyan and Andrew Zisserman. Very deep convolutional networks for large-scale image recognition. *arXiv preprint arXiv:1409.1556*, 2014.

[51] Ion Stoica, Dawn Song, Raluca Ada Popa, David Patterson, Michael W Mahoney, Randy Katz, Anthony D Joseph, Michael Jordan, Joseph M Hellerstein, Joseph E Gonzalez, et al. A berkeley view of systems challenges for AI. *arXiv preprint arXiv:1712.05855*, 2017.

[52] Christian Szegedy, Wojciech Zaremba, Ilya Sutskever, Joan Bruna, Dumitru Erhan, Ian Goodfellow, and Rob Fergus. Intriguing properties of neural networks. *arXiv preprint arXiv:1312.6199*, 2013.

[53] Tugce Tasci and Kyunghee Kim. Imagenet classification with deep convolutional neural networks, 2015.

[54] Andrei Tatar, Cristiano Giuffrida, Herbert Bos, and Kaveh Razavi. Defeating software mitigations against rowhammer: a surgical precision hammer. In *International Symposium on Research in Attacks, Intrusions, and Defenses*, pages 47–66. Springer, 2018.

[55] M. Teichmann, M. Weber, M. Zöllner, R. Cipolla, and R. Urtasun. Multinet: Real-time joint semantic reasoning for autonomous driving. In *IEEE Intelligent Vehicles Symposium*, pages 1013–1020, June 2018.

[56] Florian Tramer and Dan Boneh. Slalom: Fast, verifiable and private execution of neural networks in trusted hardware. *arXiv preprint arXiv:1806.03287*, 2018.

[57] Victor Van Der Veen, Yanick Fratantonio, Martina Lindorfer, Daniel Gruss, Clementine Maurice, Giovanni Vigna, Herbert Bos, Kaveh Razavi, and Cristiano Giuffrida. Drammer: Deterministic rowhammer attacks on mobile platforms. In *ACM Conference on Computer and Communications Security*, pages 1675–1689. ACM, 2016.

[58] S. Wang, P. Dehghanian, L. Li, and B. Wang. A machine learning approach to detection of geomagnetically induced currents in power grids. *IEEE Transactions on Industry Applications*, pages 1–1, 2019.

[59] Pete Warden. Speech commands: A dataset for limited-vocabulary speech recognition. *arXiv preprint arXiv:1804.03209*, 2018.

[60] Miloš Doroslovački Xianglin Wei, Li Li and Suresh Subramaniam. Classification of channel access attacks in wireless networks: A deep learning approach. In *40th IEEE International Conference on Distributed Computing Systems*, 2020.

[61] Han Xiao, Kashif Rasul, and Roland Vollgraf. Fashion-mnist: a novel image dataset for benchmarking machine learning algorithms. *arXiv preprint arXiv:1708.07747*, 2017.

[62] Huang Xiao, Battista Biggio, Gavin Brown, Giorgio Fumera, Claudia Eckert, and Fabio Roli. Is feature selection secure against training data poisoning? In *International Conference on Machine Learning*, pages 1689–1698, 2015.

[63] Wayne Xiong, Jasha Droppo, Xuedong Huang, Frank Seide, Mike Seltzer, Andreas Stolcke, Dong Yu, and Geoffrey Zweig. Achieving human parity in conversational speech recognition. *arXiv preprint arXiv:1610.05256*, 2016.

[64] Mengjia Yan, Christopher W. Fletcher, and Josep Torrellas. Cache telepathy: Leveraging shared resource attacks to learn DNN architectures. *CoRR*, abs/1808.04761, 2018.

[65] Fan Yao, Milos Doroslovacki, and Guru Venkataramani. Are coherence protocol states vulnerable to information leakage? In *IEEE International Symposium on High Performance Computer Architecture*, pages 168–179. IEEE, 2018.

[66] Fan Yao, Miloš Doroslovački, and Guru Venkataramani. Covert timing channels exploiting cache coherence hardware: Characterization and defense. *International Journal of Parallel Programming*, 47(4):595–620, 2019.

[67] Fan Yao, Hongyu Fang, Miloš Doroslovački, and Guru Venkataramani. Leveraging cache management hardware for practical defense against cache timing channel attacks. *IEEE Micro*, 39(4):8–16, 2019.

[68] Fan Yao, Guru Venkataramani, and Miloš Doroslovački. Covert timing channels exploiting non-uniform memory access based architectures. In *Proceedings of the on Great Lakes Symposium on VLSI 2017*, pages 155–160, 2017.

[69] Zhenlong Yuan, Yongqiang Lu, Zhaoguo Wang, and Yibo Xue. Droid-Sec: Deep learning in android malware detection. In *ACM Conference on SIGCOMM*, pages 371–372. ACM, 2014.

[70] Dongqing Zhang, Jiaolong Yang, Dongqiangzi Ye, and Gang Hua. Lq-nets: Learned quantization for highly accurate and compact deep neural networks. In *Proceedings of the European Conference on Computer Vision*, pages 365–382, 2018.

[71] Shuchang Zhou, Yuxin Wu, Zekun Ni, Xinyu Zhou, He Wen, and Yuheng Zou. Dorefa-net: Training low bitwidth convolutional neural networks with low bitwidth gradients. *arXiv preprint arXiv:1606.06160*, 2016.

## A  Model Quantization Configuration

**Weight Quantization.** Our deep learning models adopt a layer-wise *N*-bits uniform quantizer. For *m*-th layer, the quantization process from the floating-point base $\mathbf{W}_m^{\text{f}}$ to its fixed-point (signed integer) counterpart $\mathbf{W}_m$ can be described as:

$$\Delta w_m = \max(\mathbf{W}_m^{\text{f}})/(2^{N-1}-1); \quad \mathbf{W}_m^{\text{f}} \in \mathbb{R}^d \quad (3)$$

$$\mathbf{W}_m = \text{round}(\mathbf{W}_m^{\text{f}}/\Delta w_m) \cdot \Delta w_m \quad (4)$$

here *d* is the dimension of weight tensor, $\Delta w_m$ is the step size of weight quantizer. For training the quantized DNN with non-differential stair-case function (in equation 4), we use the straight-through estimator as other works [44, 71].

**Weight Encoding.** The quantized weights are represented as two's complement in computing systems. If we consider one weight element $w \in \mathbf{W}_m$, the conversion from its binary representation ($\boldsymbol{b} = [b_{N-1},...,b_0] \in \{0,1\}^N$) in two's complement can be expressed as:

$$w/\Delta w = g(\boldsymbol{b}) = -2^{N-1} \cdot b_{N-1} + \sum_{i=0}^{N-2} 2^i \cdot b_i \quad (5)$$

We perform weight quantization during the training for all the models except 5 architectures listed in Table 1 using ImageNet dataset. For ImageNet architectures, we use post-quantization on the pre-trained model.

## B  DNN Architecture Configuration

For MNIST classification we use simple LeNet [31] architecture with two convolution layers and two fully-connected layers. For, VGG-13 and VGG-11 we use conventional architecture delineated as shown in [50]. Each of the architecture features three fully-connected layers and remaining convolution networks. Our AlexNet architecture contains five sets of convolution layers, ReLu and Maxpooling followed by three dropout and fully-connected layers [53]. Finally, for ImageNet, we leverage the PyTorch official website's trained models in torch vision. For this reason, we quantize the network after training for ImageNet simulation (i.e.,post quantization).

## C  DNN Training Configuration

Our training platform is in PyTorch using GPU 1080 Ti platform. For MNIST dataset training, we have the following configuration: batch size 256, learning rate 0.1 , momentum 0.9, weight decay $3e^{-4}$ and SGD optimizer with gamma 0.1. For CIFAR-10 our training configuration is as follows: learning rate 0.1, optimizer SGD with 0.1 gamma and momentum 0.9, batch size 128, training epoch 200 and weight decay $3e^{-4}$. For the speech command dataset, we only train the network for 70 epoch with a $1e^{-2}$ weight decay, $1e^{-4}$ learning rate and SGD optimizer with a batch size of 128. For ImageNet dataset, we directly run the attack on a pre-trained model. We perform post quantization on a pre-trained torch vision model.

## D  Targeted Bit-flip Chain for DNN Models

In Table 4, we list the bit chains identified by DeepHammer. For networks without any residual connections (i.e, VGG and AlexNet), we observe that most of the bit-flips are located at the earlier layers, indicating that bit-flip perturbation in the weight accumulates as it passes through later layers. However, for ResNet architectures, most bit flips are identified both at the front and the end of the network. As a result, we conclude network topology may affect the locations of the vulnerable model weight bits.

| Dataset | Architecture | Chain of bits (page#, bop, mode) |
|---|---|---|
| F-MNIST | **LeNet** | (1,1519,0)→(4,12595,0)→(159,302,1) |
| Speech | **VGG-11** | (6859,23008,1)→(1,1519,1)→(125,799,0)→(6866,23008,0)→(2533,20816,0) |
| Speech | **VGG-13** | (1,2007,0)→(6904,25856,1)→(5465,2704,1)→(2155,6424,0)→<br>(1557,48,0)→(2778,15896,1)→(6914,25856,1) |
| CIFAR-10 | **ResNet-20** | (66,14055,0)→(4,25639,0)→(1,24399,0)→(9,16175,0)→(5,25047,0)→<br>(2,29095,0)→(3,32759,0)→(10,9735,0)→(13,9031,0)→(14,25423,0)→<br>(55,22071,0)→(27,22071,0)→(50,15431,0)→(63,21071,0)→(21,25127,0)→<br>(12,23863,0)→(18,2215,0)→(39,21935,0)→(45,18655,0)→(48,21047,0)→(51,28719,0) |
| CIFAR-10 | **AlexNet** | (1,4319,0)→(21,4991,0)→(48,32135,0)→(355,1943,0)→(483,11487,0) |
| CIFAR-10 | **VGG-11** | (591,7848,0)→(316,16407,0)→(111,26153,0) |
| ImageNet | **MobileNet-V2** | (1,30855,0)→(2,3399,1) |
| ImageNet | **SqueezeNet** | (23,5167,1)→(7,11895,1)→(12,783,0)→(4,30071,0)→(21,26967,0)→<br>(6,1671,0)→(142,3062,0)→(10,12343,0)→(9,13847,0)→(8,1087,1)<br>(304,23550,0)→(24,13423,1)→(5,631,0)→(141,10351,0)→<br>(60,19615,0)→(37,15231,0)→(94,4215,0)→(139,28959,0) |
| ImageNet | **ResNet-18** | (1,29287,1)→(2,26855,0)→(95,2967,1)→(29,1855,1)→(93,15943,0)→<br>(9,1167,0)→(22,21791,0)→(31,14535,0)→(1571,16296,0)→(60,25367,0)→<br>(106,28031,0)→(13,18191,0)→(201,30055,0)→(384,30311,0)→<br>(134,24983,0)→(52,17543,0)→(2144,13568,0)→(1731,17648,1)→(565,1464,0)→<br>(268,26823,0)→(45,7295,1)→(931,31968,0)→(9321,7768,0)→(224,22887,0) |
| ImageNet | **ResNet-34** | (112,5111,0)→(39,3103,0)→(90,23831,0)→(11,1567,0)→(21,4503,0)→(57,983,0)→<br>(278,7511,0)→(63,1967,0)→(203,4407,0)→(236,20471,0)→(164,23711,0)→(550,30648,0)→<br>(42,21911,1)→(46,29103,1)→(40,27575,1)→(47,14743,0)→(547,2998,1)→(433,23175,0)→<br>(26,11647,0)→(66,5015,0)→(798,31536,0)→(111,15863,1)→(28,24495,0) |
| ImageNet | **ResNet-50** | (20,17911,0)→(62,31870,1)→(118,9342,1)→(16,17503,1)→(60,13438,1)→(379,14207,0)→<br>(115,23678,1)→(54,17719,0)→(100,25807,0)→(88,19599,0)→(37,17647,0)→(2179,24568,0)→<br>(2824,14432,0)→(5,31079,0)→(99,16231,0)→(82,13439,0)→(225,10111,0)→(40,7295,1)→<br>(4,8967,0)→(4757,8592,0)→(9,2455,0)→(2905,22624,0)→(2109,31432,0) |

Table 4: Illustrations of identified shortest chains of targeted bits for all DNN models flip-aware bit search algorithm.